\documentclass[onecolumn]{aa} % for a paper on 1 column  
\usepackage{graphicx}
\usepackage{txfonts}
\begin{document}
   \title{Spectral analysis of LMC X--2 with XMM/Newton: unveiling the emission process in the extragalactic Z-source. }

   \author{G. Lavagetto
%          \inst{1}
          \and
          R. Iaria
%          \inst{1}
          \and
          A. D'A\`\i\,
%          \inst{1}
          \and
          T. Di Salvo
%          \inst{1}
          \and N.R. Robba
%          \inst{1}
          }

   \offprints{G. Lavagetto}

   \institute{Dipartimento di Scienze Fisiche ed Astronomiche (DSFA), Universit\`a degli studi di Palermo, via Archirafi 36, 90123 Palermo (PA) Italy
     \email{giuseppe.lavagetto@fisica.unipa.it}\\
             }
\titlerunning{Spectral analysis of LMC X--2}
%\authorrunning{Lavagetto et al.}
   \date{Received 2007; accepted }
\abstract
%context
{}
%aims
{We present the results of the analysis of an archival observation of LMC X--2 
performed with XMM/Newton. The spectra taken by high-precision instruments have never been analyzed before.}
%methods
{We find an X--ray position
for the source that is inconsistent with the one obtained by ROSAT,
but in agreement with the Einstein position and that of the optical counterpart. The
correlated  
spectral and timing behaviour of the
source suggests that the source is probably in the normal branch of
its X-ray color-color diagram. The spectrum of the source can be
fitted with a blackbody with 
a temperature 1.5 keV plus a disk blackbody at 0.8 keV. Photoelectric
absorption from neutral matter has an equivalent hydrogen column of $4 \times
10^{20}~\mathrm{cm}^{-2}$. An emission line, which we identify as the O VIII
Lyman-$\alpha$ line, is detected, while no feature due to iron is
detected in the spectrum.}
%results
{We argue that the emission of this source
can be straightforwardly interpreted as a sum of the emission from
a boundary layer between the NS and the disc and a blackbody component coming
from the disc itself. Other canonical models that are used to fit
Z-sources do not give a satisfactory fit to the data. The
detection of the O VIII emission 
line (and the lack of detection of lines in the iron region) can be
due to the fact that the source lies in the Large Magellanic Cloud.
}
%conclusions
{}
\keywords{}
\maketitle

\section{Introduction}
Low  Mass   X-ray  Binaries  (LMXBs)   are  binary  systems
harboring a compact object accreting mass from a late-type  star. 
Systems hosting a weakly magnetized neutron star (NS) are  usually divided 
into  two classes: the Z  sources, with  luminosities close  to  the Eddington
luminosity, $L_{\rm  edd}$, and  Atoll sources,  usually with  lower 
luminosities  of $\sim  0.01-0.1\ L_{\rm edd}$.  This classification 
relies  upon the correlated X-ray spectral and timing properties, 
namely the  pattern traced out by individual sources
in the X-ray  color-color diagram  (CD,  Hasinger \&  van der  Klis 1989).  
The seven  known  (Galactic) Z sources usually describe  a complete Z-track  
in the  CD on timescales  of a  few days,
while Atoll sources cover their pattern on the CD on a longer
timescale (usually several weeks).

The correlated spectral and timing variability of LMC X--2 has
suggested its classification as a Z-source (Smale \& Kuulkers 2000). An extensive study of
these characteristics, using data taken with the PCA on board RXTE, 
has allowed to observe the complete Z-track in the CD which is completed
in about 1 day (Smale, Homan \& Kuulkers 2003). Smale \& Kuulkers (2000) found
evidence of a periodicity of $8.16~\mathrm{h}$ which they tentatively ascribe to an orbital period. A similar orbital period modulation has been found by Cornelisse et al. (2007) in the VLT optical lightcurve of the companion. From the shape of the 
Z-pattern described in the CD,
 it was inferred that LMC X-2 falls squarely into the
category of the 
so-called ``Sco-like'' Z-sources due to its common characteristics
with Sco X-1 (Smale et al. 2003). Its very high luminosity ($\sim
0.5-2 L_\mathrm{edd}$; Markert \& Clark 1975; Johnston, Bradt \& Doxsey
1979; Long, Helfand \& Grabelsky 1981; Bonnet-Bidaud et al. 1989; Christian \& Swank 1997;
Smale \& Kuulkers 2000) makes it the brightest LMXB 
known together with Sco X--1. There are other analogies with Sco
X--1: the optical counterpart is a faint, $M_V\sim 18.8$, blue star 
(similar to the optical counterpart of Sco X--1), and both sources show a similar
correlation between the optical and the X-ray lightcurve during flares
(McGowan et al. 2003). In some ways, we can consider LMC
X--2 an extragalactic twin to Sco X--1.  Measurement of Doppler effect on emission lines in the Bowen region allowed Cornelisse et al. (2007) to constrain the mass ratio of LMC X-2 to be $\le 0.4$.

However, there is no recent X-ray spectral study of this source:
a spectral analysis has been carried out by Bonnet-Bidaud et
al. (1989) using EXOSAT data, where the spectrum of the source was
fitted with a model consisting of a blackbody $kT\sim 1.3~\mathrm{keV}$ plus a thermal
bremsstrahlung ($kT\sim 5~\mathrm{keV}$), or alternatively with a comptonized thermal model with $kT\sim 3~\mathrm{keV}$. They found no detectable feature in the Fe K-$\alpha$ range. In their survey of Low Mass X-Ray Binaries with Einstein, Christian \& Swank (1997) report that a fit in the 1-20 keV band with an unsaturated Comptonized model (i.e. a cutoff power-law) gave a reasonable fit with $\Gamma = 1.4$ and $kT = 6.4~\mathrm{keV}$.
Schulz
(1999), using ROSAT data in the 0.1--2.4 keV band, fitted the data using a blackbody with a temperature $kT = 1.5~\mathrm{keV}$ plus a
thermal bresstrahlung model with a temperature of $kT=5~\mathrm{keV}$. Also, a
gaussian emission line at 0.9 keV was evident in the spectrum. The
accuracy of this study is obviously limited by the narrow band and the
relatively poor spectral resolution of ROSAT. Smale \& Kuulkers (2000) studying
RXTE/PCA data of the source found that the data can be well described
by a cutoff power-law (e.g. a completely Comptonized component) with a
cutoff temperature of $2.8~\mathrm{keV}$, or by a blackbody plus
bremsstrahlung model with $kT_\mathrm{bb}\sim 1.5$ keV, $kT_\mathrm{brems}
\sim 4.5$ keV. In this case the blackbody accounts for about 20\% of the
total emission: as the authors point out, however, this model is not
physically realistic given the enormous emitting volumes required for
the bremsstrahlung emission.

These low-resolution spectra differ noticeably from the  X-ray
spectra  of other Z-sources, which are usually described in terms of a 
two-component model. The spectral models of Z-sources usually follow
one of two main paradigms: the Eastern model (Mitsuda et al. 1989), in which the spectrum of
the source is interpreted as a sum of an emission coming directly from
the compact object (Comptonized or not) plus a
blackbody emission from the disc, and the Western model (White et al. 1986), where the
emission is due to a hot blackbody coming from the central source, plus a
Comptonized component that is due to the emission from an accretion
disc corona surrounding the disc (no direct thermal emission from the disc is
seen as photons emitted there are reprocessed and Comptonized in the
corona).

Broad emission  lines (FWHM  up to  $\sim 1$ keV)  at energies  in the
range 6.4 -- 6.7 keV have been observed in the spectra of all galactic
Z-sources, with the noticeable exception of GX~5-1 (Asai et al. 1994). These
lines are identified with  the K$\alpha$ radiative transitions of iron
at different  ionization states. Sometimes
an iron  absorption edge at energies  $\sim 8$ keV  has been detected (see e.g. Di Salvo et al. 2001).

In this paper we analyze an archival XMM/Newton observation of LMC
X--2, which gives us the opportunity to perform the first high
resolution spectral study available to date on this source.

\section{Observation and data analysis}
\label{sec:obs}

XMM/Newton (Jansen et al. 2001) has observed LMC~X-2 between
2003-04-21, 20.21:03 and 2003-04-22, 03:03:58. The source
was observed in High event rate mode with both the Reflection Grating
Spectrometers (RGS1 and 2, den Herder et al. 2001) for a common exposure time of $\sim 19.5$ ks.
 The EPIC-PN camera (Str\"uder et al. 2001) observed the source in Small Window
mode with Medium filter, for a total exposure of 11.2 ks. Of the two
MOS, only MOS2 data in Fast Uncompressed mode are available, for a mere
457 s of exposure.

 The OM data are available and an optical counterpart candidate is detected
 in all three bands (UVW1,UVW2, UVM2), from which we can deduce the
 instrumental magnitudes. However, recent VLT data have been observed by
 Cornelisse et al. (2007) and surely gave more interesting information than
 what could be achieved with the OM; we, therefore, do not consider the OM data in this paper any further as it
would not contribute significantly to our scientific results.

The raw, full bandpass average count rates are 3.7, 4.6 cts/s for RGS1 and RGS2 (0.2-2.1 keV), and 120 cts/s for Epic-PN (0.15-15.0 keV). This means
that, given the use of the small window mode, any pileup issue is avoided for this source.
%We analyzed RGS and PN data, extracting source information using
%standard techniques for image, spectrum and lightcurve extraction for
%point-like sources.
In all of our data reduction and analysis we used the latest available
software packages, i.e. SAS version 7.0, Heasoft 6.2 and Xspec 11.3.2ad. 

\subsection{X-ray position}
\begin{figure}[h]
  \centering
  \includegraphics[angle=90,scale=0.4]{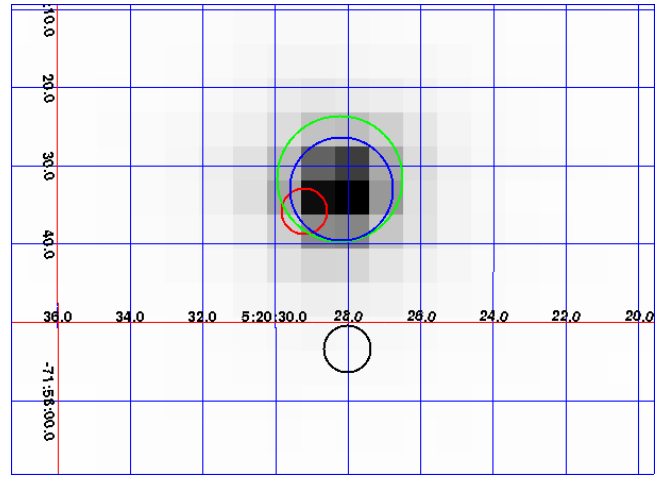}
  \caption{EPIC-PN Image of LMC X-2, toghether with the XMM/Newton X-ray
    position (blue circle), the Einstein X-ray position (green
    circle), the position of the optical couterpart of LMC~X--2 (red circle), and the position of the X-ray source according to the ROSAT All Sky Survey (black circle). The circles represent the derived  error in the position. The coordinate grid refers to the J2000 equinox.}
  \label{fig:imaging}
\end{figure}

We extracted the PN image of the source (see figure \ref{fig:imaging}) using the 
standard techniques described by the SAS online
documentation\footnote{Available at \texttt{http://xmm.esac.esa.int/sas/7.0.0/documentation/threads/}}: first of all, we looked for strong particle background intervals in the data, and found that there is no contamination to the data. Next, we extracted the image of the source using the xmmselect program that is part of the SAS distribution. The
EPIC-pn image allowed us to determine the coordinates of the source (J2000)
$\alpha=5^h 20^m 28.2^s,\delta=-71^h 57^m 33^s$. While this position
(given the error on the position obtained from the EPIC-pn image, $\sim 6.6$ arcsec or 1.5 pixels in the image, coincident with the nominal image resolution of the instrument) is  
in agreement with the one of the optical counterpart (Pakull, 1978),
and with the X-ray position obtained with Einstein (Long et al., 1981), 
it differs significantly ($20.3$ arcseconds)
from the ROSAT position (Fuhrmeister \& Schmitt, 2003). We can
therefore assume that the position we determined is the more accurate
X-ray position actually available.
\subsection{Timing properties}

\begin{figure}[h]
  \centering
  \includegraphics[scale=1]{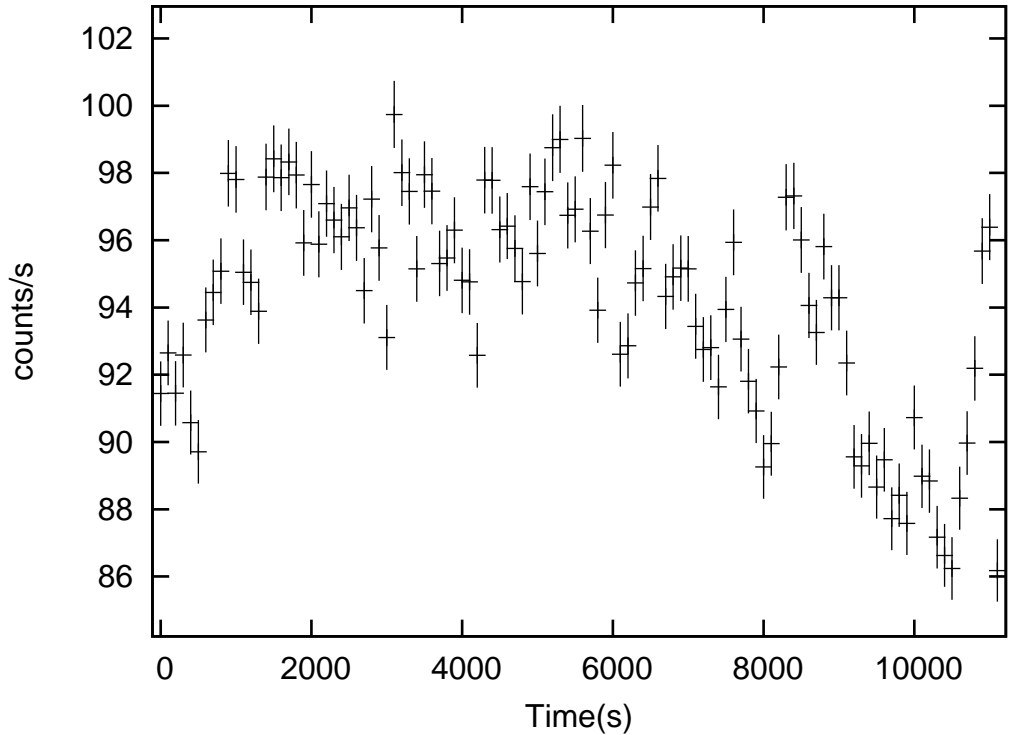}
  \caption{PN light curve of LMC~X-2. The bin time is 100 s. We plot
    the count rate versus time (in seconds) from the beginning of the observation. }
  \label{fig:lcurve}
\end{figure}

\begin{figure}[h]
  \centering
  \includegraphics[scale=1]{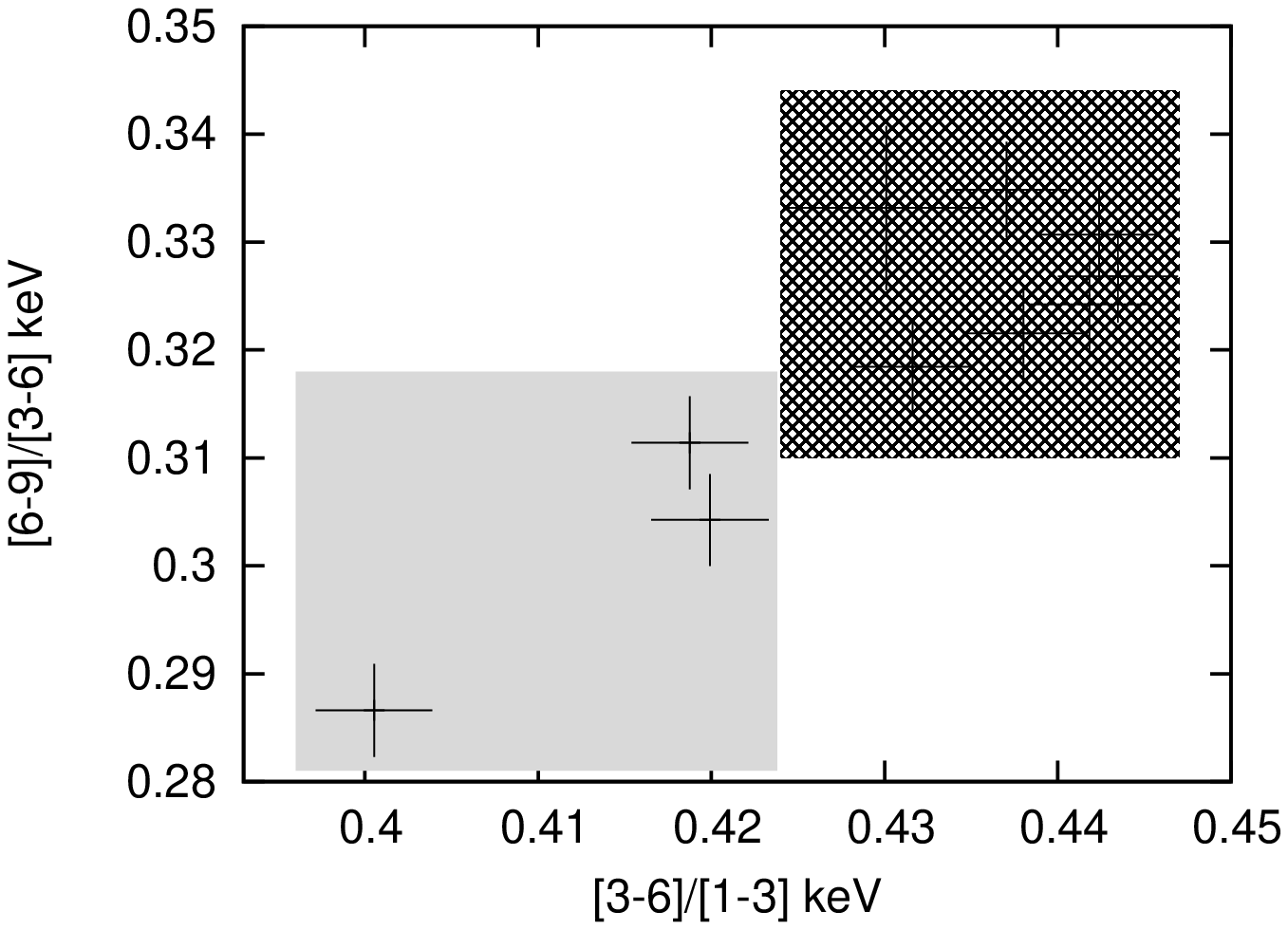}
  \caption{Color-Color diagram of LMC X-2 extracted from PN data. The bin size for each point is 1.2 ks, the error bars are at $90\%$ confidence. With time, the source moves from the bottom left to the top right of the diagram.}
  \label{fig:ccdia}
\end{figure}

Following the standard techniques, we extracted the
background-subtracted lightcurve for the EPIC-PN:
 even if taken at different bin scales, 
the PN lightcurve, although it shows some variability (in particular in the final part of the observation), this variability is not as strong as one would expect if the source was in the so-called flaring branch, as can be seen from figure \ref{fig:lcurve}.
Interestingly enough, while the count rate remains almost the same
during the whole observation, the PN color-color diagram shows a slight
drift during the observation towards higher values of both the soft
and the hard color (defined as the ratio of the count rate in the
range 3--6 keV to the count rate in the range 1--3 keV, and as the
ratio  the count rate in the range 6--9 keV to the count rate in the
range 3--6 keV, respectively), as can be seen from figure \ref{fig:ccdia}.

\begin{figure}[h]
  \centering
  \includegraphics{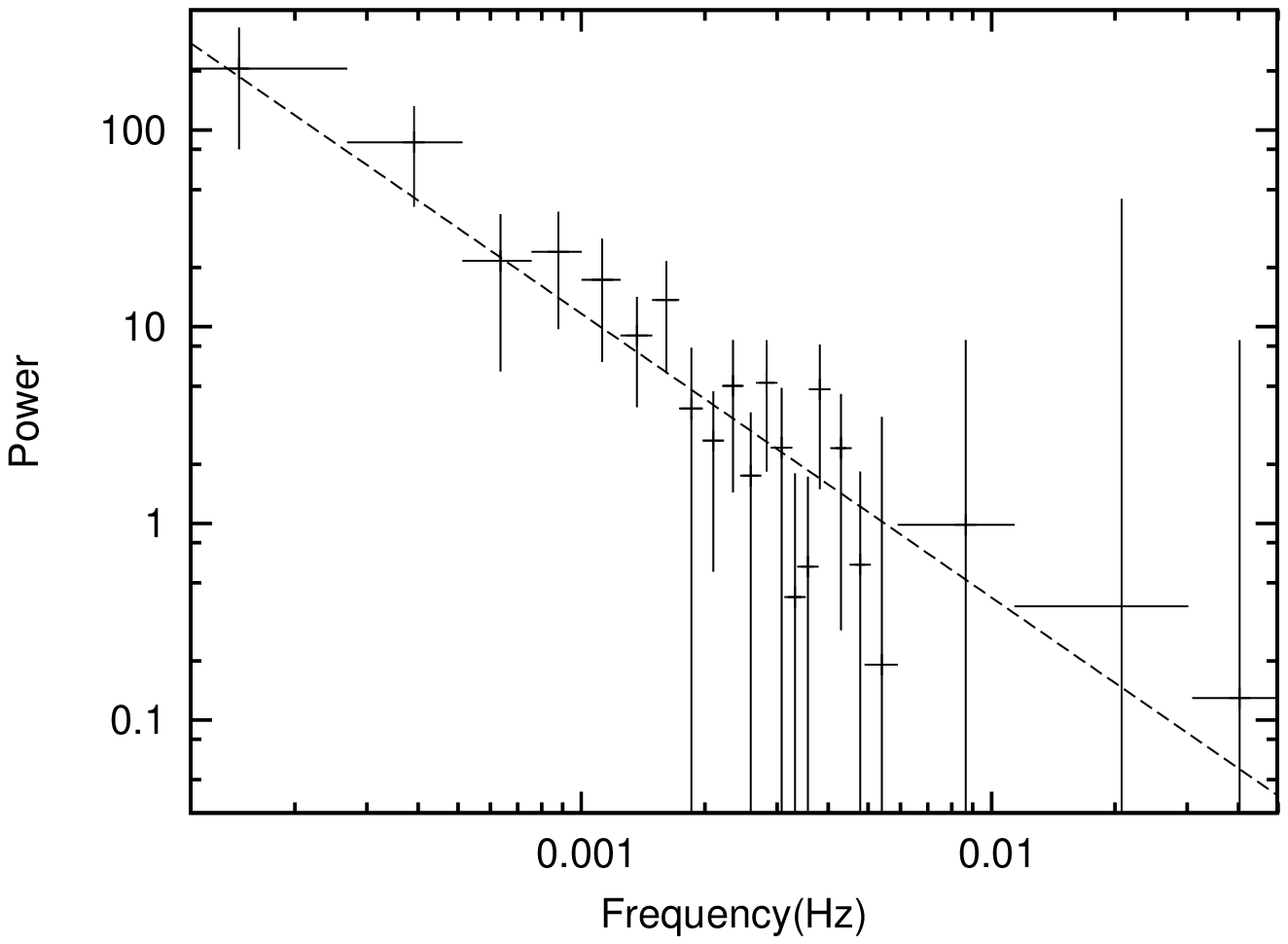}
  \caption{White noise subtracted power density spectrum of LMC
    X--2. The dashed line represents the best fit power-law
    $\nu^{-\gamma}$ with $\gamma=1.4$.}
  \label{fig:psd}
\end{figure}

Although the statistics is poor, we were able to extract a
low-frequency power spectrum in order to obtain information on the
very low frequency noise. We fitted the white noise-subtracted power
spectrum with a power law $P(\nu) \propto \nu^{-\gamma}$ and found
$\gamma = 1.4 \pm 0.3$, $\mathrm{rms} (0.001-1~\mathrm{Hz})=2.2\pm0.4\%$ with $\chi^2/d.o.f. = 42/100$  (see figure \ref{fig:psd}). Given the value of
$\gamma$, which is significantly higher than 0.6, the value found by Smale et
al. (2003) in the Horizontal Branch, and the drift of the source in the 
CD, we can affirm that the source is either in the Normal or the
Flaring Branch, as confirmed by the relatively high luminosity state
(see below).  Moreover, the relatively constant count-rate (see figure \ref{fig:lcurve}) suggests that 
the source is not in the Flaring Branch, as in that state frequent and 
intense flares are normally seen (Smale et al. 2003). This is also confirmed by the value of rms, which is compatible with what found by 
Smale et al. (2003) in the Normal Branch or in the Horizontal Branch, but not with the higher value ($3.2\pm 0.1 \%$) found in the Flaring Branch. 

\subsection{Spectral fitting}
In order to verify if the drift in the CD affected
significantly the continuum model, 
we divided the observation into two parts according to its
movement along the color-color diagram (see the shadowed portion of figure \ref{fig:ccdia}, and extracted spectra for the
two zones separately, in order to fit toghether the
most spectrally homogeneous data. In extracting the spectra we
followed again the standard SAS recipe for point-like
sources. 
 However, we found that the two PN spectra could be fit
together with little discrepancy, and no statistically
significant variation was found between the spectral fits of the two
zones. This means that the spectral variability of the source is not
accentuated enough to introduce sensible differences.

Given this result, we decided to extract a single spectrum from the
whole dataset. Therefore we run the standard extraction tool software
again and accumulated background subtracted spectra for both the PN and the
two Reflection Grating Spectrometers\footnote{We discarded MOS data
  given the very short exposure.}. In extracting RGS spectra, we
kept into account the exact source position we determined from the PN
image analysis.

In the
PN, we verified the absence of flaring particle background extracting an 
high-energy (10-12 keV) lightcurve, where the source shows a steady count rate of 
4 count/s, showing no particular flaring activity; this means that any 
particle background is not important for this high-flux source.
 We selected photons coming from a circular region of $20''$
centered on the brightest pixel (the source coordinates), and for the
background we selected
photons from a region of the same shape and area in a blank sky region
near the source, with the
center on the same detector row as the source area. For RGS 1 and 2
the spectra were extracted using the standard pipeline, indicating the
our newly refined position. All spectra 
were rebinned in order to have at least 25 counts per channel, so that we are allowed to use the $\chi^2$ statistics.

\begin{figure}[h]
  \centering
  \includegraphics[]{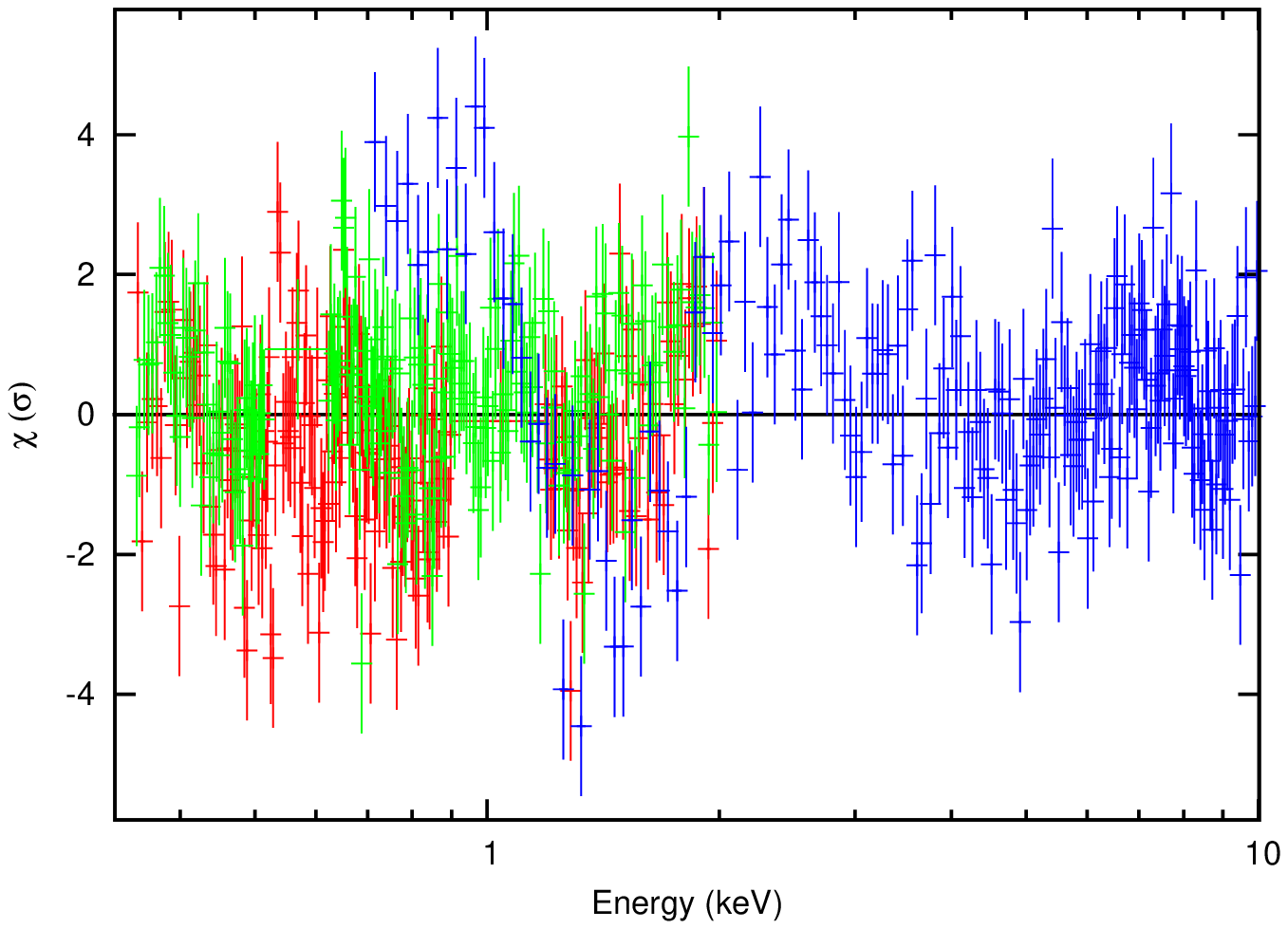}
  \caption{Residuals with respect to the blackbody+bremsstrahlung model. The
    mismatch between the two RGS (red and green) and the EPIC-pn
    (blue) data below 1.3 keV is evident.}
  \label{fig:bremsres0}
\end{figure}

We decided to use RGS data in the standard energy range $0.35-2.0$ keV, and PN
data in the $0.6-10$ keV energy range. We fitted these spectra to a
two-component model constituted by a blackbody  plus a bremsstrahlung
component, which is the only two component model which gave a
satisfactory fit to EXOSAT (Bonnet-Bidaud et al. 1989), ROSAT (Schulz 1999) and RXTE data (Smale \& Kuulkers 2000)\footnote{Einstein data were not fitted to such a model (Christian \& Swank 1997), but only to a cutoff power-law model. Such a model is discussed below.}.
  Both components were modified by photo-electric absorption from
  neutral matter, and a normalization  
constant was introduced in order to account for 
uncertainties in the cross-calibration between instruments. We found
that the model gave a fit that, although not rejectable
statistically ($\chi^2/d.o.f.= 5916/5529$), is not satisfactory 
if one looks at its
residuals: the PN residuals clearly show sign of a poor fit of the
continuum, as they jiggle around the zero residual line (see figure
\ref{fig:bremsres0}).

In particular, these residuals show that there is a strong mismatch between the
two RGSs and the EPIC PN below 1.3 keV. This is probably due to
calibration issues of the PN for bright sources (see e.g. Boirin \& Parmar 2003). We therefore decided to use PN data limiting ourselves to
the 1.3-10.0 keV range.

Using this reduced spectral range for the PN, we found that the data could be
fitted well with this two-component model: we found a blackbody
temperature of $1.4$ keV and a bremssstrahlung temperature of $7.7$ keV (see Table 1 for detailed results of all fits).
The resulting blackbody radius is $11.6\pm0.3$ km\footnote{All quantities through this section are derived for a distance of the source of 50 kpc.}, while the radius of the
bremsstrahlung emitting region is $\sim 10^8$~km; the blackbody accounts
for about $30\%$ of the total luminosity. We found
evidence in the residuals for an emission line at  $0.653\pm 0.002$
keV, which we identify with a Ly-$\alpha$ emission line from O
VIII with an equivalent width of 2.5 eV (see figure 7, table 1 for
details). Although 
the above described two-component model gives a 
satisfactory fit to the data, it is
not physically reasonable: first of all, the emission region for the 
bremsstrahlung component is  presumably orders of magnitude larger than the
whole system, which is, assuming NS mass of 2 M$_\odot$ and a mass ratio of 0.4 (Cornelisse et al 2007), in the order of $10^6$ km. Moreover most of the emission should come from that component, while 
one would expect that at least half of the emission would come from the direct 
emission from the compact object (e.g. the blackbody). We searched
therefore for an alternative model to fit the data.

We firstly fitted the spectra to a simple model made of a comptonized
component: Bonnet-Bidaud et al. (1989), Chiristian \& Swank (1997) and Smale \& Kuulkers (2000) used a 
simple cutoff powerlaw model to fit  data of the source. Using such a simple 
model to fit our data, however, proved to be impossible since the power-law 
index we obtain is below 1 ,so that this powerlaw does not describe a 
Comptonized component: 
$\Gamma <1$ implies $y=4kT/m_e c^2<0$ which is impossible. Thus we tried to fit the data to a more complex comptonized model,
described by \textit{compTT} model inside xspec,
plus photo-electric absorption from neutral matter and the O
VIII Lyman-$\alpha$ emission line. This model is statistically as good as the preceding one (the $\chi^2/d.o.f.$ is 5571/5404), but the parameters of the Comptonized component are
quite unusual if compared to what found in other Z-sources: we find a
seed photons temperature between 10 and 50 eV, an electron temperature
of $1.73\pm 0.05$ keV and an optical depth value of $\tau = 23.1\pm 0.2$, while normally
in Z-sources the seed photons (and the Comptonizing electrons) are
hotter, and the optical depth is of the order of 10. The Wien's radius of
the seed photons, although poorly constrained, is in the order of
10$^4$ km. This would suggest that the emission we see comes from an
extended corona, while direct emission from the compact object is
completely shadowed. 

The single-Comptonized component model resembles somewhat to the ``Birmingham model'', where the emission of
the central source is described by a blackbody and the rest of the
emission comes from an accretion disc corona. This model has recently been applied to Z-sources (Church et al. 2006). We then tried to fit the data to a two-component model constituted by
a blackbody and a cutoff powerlaw, to test if the 
Birmingham model was also acceptable. We found that, although this 
model could fit well the data, the cutoff powerlaw index is again below 1,
so that this powerlaw does not describe a Comptonized component. If we try
to fix the power-law index to values common in fits of spectra of other 
Z-sources, like 1.9 (Sco X--1, Barnard et al. 2003) or 1.7 (GX~340+0,
Church et al. 2006), the fit we obtain
to the data is unstable, as the high energy cutoff becomes larger than
15 keV and is unconstrainable; moreover, the residuals jiggle
significantly around zero and the corresponding
$\chi^2/d.o.f. =5844/5404$, is very high compared to the one found with
other models.

We found that data could also be fitted by 
a conventional ``Eastern model'', which is often used to describe the
X-ray spectra of Z-sources. In particular, we fitted the data with a two-component
model constituted by a
blackbody and a disk blackbody, modified by photo-electric absorption from 
neutral matter and the O VIII Ly-$\alpha$ emission line. This model
is again not statistically worse than the former two (we obtain $\chi^2(d.o.f.) = 5555(5404)$). In this
case, we find a blackbody temperature of $1.54$ keV, a radius of $14.9 \pm 0.02$
km and a luminosity of $1.6 \times 10^{38}~\mathrm{erg~s^{-1}}$\footnote{In the whole paper, luminosities are calculated in the 0.1-10 keV energy range, for a distance of 50 kpc.}, while
the disk emission accounts for the remaining $33\%$ ($8 \times 10^{37}~\mathrm{erg~s^{-1}}$) of the emission
with an inner disk temperature of $0.815$ keV and a lower limit on the inner radius of 27.5 km. Full results, toghether with corresponding errors are reported in Table 1.

No residuals could be noticed in the iron K-$\alpha$ region, confirming the result obtained by Bonnet-Bidaud et al. (1989). Since feature in that spectral range are usually present in Z-sources, it could well be possible, in
line of principle, that the short exposure time did not allow to
accumulate the statistics required to show a significant iron line. To
test this, we added to the best-fit models a gaussian emission line
with a fixed energy of 6.7 keV and $\sigma$ of 200 eV(which is quite
typical of lines found in Z-sources). We found that, considering the
upper limit on the line flux (at 
90\% confidence), the equivalent width of the line is $\le 20$ eV,
which is about half of the lowest equivalent width found, for example,
in BeppoSAX observations of Cyg~X--2 (Di Salvo et al. 2002).

The total luminosity ($0.3-10.0~\mathrm{keV}$) of the source is, for
 all the models, $\sim 2.3\times 10^{38}~\mathrm{erg~s^{-1}}$, that is
 at least $1.3$ times the
Eddington luminosity for a $1.35~M_\odot$ NS (we assumed a pure hydrogen accreting column, which is a reasonable assumption given the low metallicity of stars in the LMC). In all our calculations
we assumed a distance of 50 kpc (Feast 1999).
% is unphysical again: the seed photon radius is very large,
% although poorly constrained. It is then hard to believe that the total
% luminosity we get (about 1-2 times the Eddington luminosity) can come
% from a region that far from the central compact object. Moreover, to
% justify the fact that we do not see the central accretion region, we
% should suppose that the system has an high inclination angle. But this
% contrasts with the finding by Kuulkers et al (????) that Sco-like
% Z-sources are systems which are most likely face on (i.e. they have
% little or no inclination).

\begin{figure}[h]
   \centering
   \includegraphics[angle=270,scale=0.5]{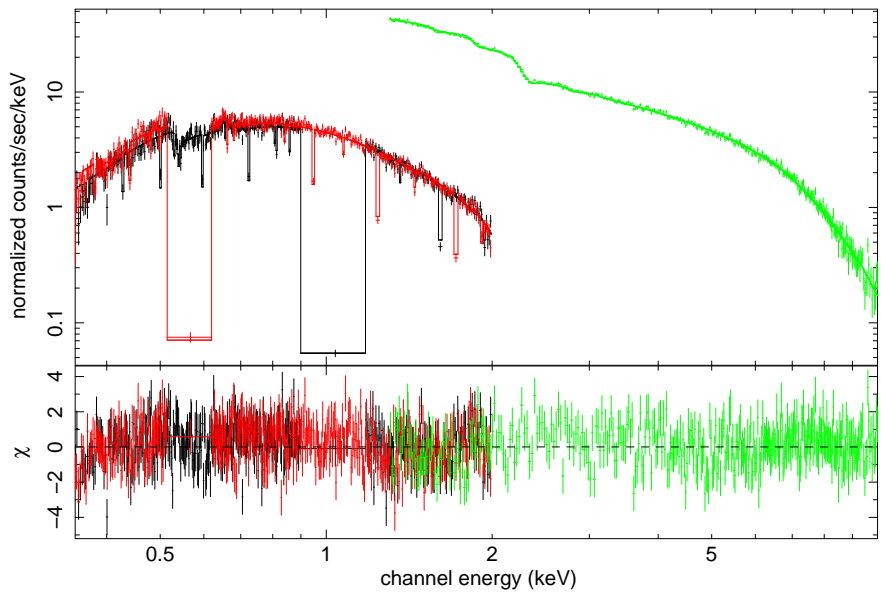}
   \caption{Data with best fit model, consisting of a blackbody and a
     disk blackbody and relative residuals in units of $\sigma$ for RGS1, RGS2 and PN, respectively. The drops in the RGS1 and RGS2 count rates are due to failing CCDs.}
   \label{fig:fitres1}
 \end{figure}

\begin{table}[h]
  \centering
  \begin{tabular}[c]{|l|c|c|c|c|}
    \hline
            Model                    & blackbody +  &
            comptt & blackbody +  & Birmingham\\
 &bremsstrahlung &  &disk blackbody & Model\\
\hline
N$_\mathrm{H}$ (10$^{20}$ cm$^{-2}$) & $8.35\pm 0.12$ &$10.00 \pm{0.12}$ &$4.04^{+0.11}_{-0.12}$&$6.67\pm{+0.12}$ \\
kT$_\mathrm{bb}$ (keV)     & $1.411\pm 0.016$ & -& $1.543\pm 0.009$& $1.66 \pm 0.02$\\
R$_\mathrm{bb}$ (Km)       &$11.6 \pm 0.3$ & -&$14.9 \pm 0.02$ &$10.1^{+1.2}_{-0.2}$\\
L$_\mathrm{bb}$ (erg/s)    &$0.67 \times 10^{38}$& -& $1.6 \times 10^{38}$ & $0.96\times 10^{38}$ \\
kT$_\mathrm{w}$ (keV)      &- & $0.037^{0.013}_{0.027}$& -& -\\
kT$_\mathrm{2}$ (keV)      &7.72$\pm 0.012$ & $1.73\pm 0.05$& $0.815\pm 0.002$&$3.18\pm 0.04$ \\
$\tau$                   &- & 23.1$\pm 0.2$& - & -\\
R$_\mathrm{w}$ (km)        &- & $(11 \pm 9) \times 10^3$ &- & -\\
R$_\mathrm{br}$ (km)       & 3.7e+8 & - & - & -\\
R$_\mathrm{diskbb}$ (Km)       &- & -&$27.52 \pm 0.05$ & -\\
$\gamma$                  &- & - &- & $0.963\pm 0.007$\\
PL norm                   &- & - &- & $ 0.1047\pm 0.0004$\\
L$_\mathrm{tot}$ (erg/s)   &$2.3 \times 10^{38}$&$2.3 \times 10^{38}$ &$2.24 \times 10^{38}$ &$2.26 \times 10^{38}$ \\
E$_\mathrm{line}$ (eV)    &$653\pm 2$  & $653\pm 2$& $653\pm 2$ & $653\pm 2$\\
$\sigma$(eV)              &5(fixed)  &5(fixed) &5(fixed) &5(fixed) \\
Eqw (eV)                  &2.53  &2.61 &  2.9 & 2.86\\
$\chi^2_\mathrm{red}$ (d.o.f.)&  5534 (5404) & 5571(5404) & 5555 (5404)& 5485 (5403)\\ 
\hline
  \end{tabular}
  \caption{Results of the fits to LMC~X--2 data with the four spectral
    Models. Uncertainties are at 90\% confidence level for a single parameter.
{\footnotesize $kT_\mathrm{bb}, R_\mathrm{bb}$ and L$_\mathrm{bb}$ are
the blackbody temperature, radius of the emitting region and
Luminosity. $kT_\mathrm{2}$ is the temperature of the second thermal
component, that is the bremsstrahlung temperature, the electron
temperature, the disk blackbody inner temperature, and the cutoff
temperature, respectively. $\tau, R_\mathrm{w}$ are the optical depth
and the Wien radius of the seed photons of the compTT
model. R$_\mathrm{br}$ is the radius of the bremsstrahlung emitting
region, R$_\mathrm{diskbb}$ is the radius of the disk blackbody,
$\gamma$ is the powerlaw photon index and the normalization of the
powerlaw is  in units of  photons keV$^{-1}$ cm$^{-2}$ s$^{-1}$ at 1 keV, Eqw is the equivalent width of the gaussian emission lines, $L_\mathrm{tot}$ is the total luminosity of the model in the 0.1-10 keV energy range. All quantities are calculated assuming a distance of 50 kpc (Feast 1999).}}
  \label{tab:fit}
\end{table}
As can be seen from figure \ref{fig:fitres1}, using the
``eastern-like'' model the fitting is good
enough and no other significant local feature is present in the
spectra.

\section{Discussion}

We have fitted the spectrum of LMC~X--2 taken with the EPIC-PN and RGS
instruments on board  XMM/Newton with four
different two-component models which describe completely different physical
processes. Since they are all statistically acceptable, in order to
choose a best-fit model we should inspect which one looks as the most
physically reasonable. 

The first model is constituted by blackbody plus a bremsstrahlung
component. This model is clearly unphysical as the bremsstrahlung
emission region is $3.6\times 10^8$ km, which is more than one order
of magnitude larger than the orbital separation for a system with an
orbital period of 30 ks. We can therefore safely discard this model
without any further discussion.

The second model involves a single, thick Comptonized component with a
large Wien radius of the emitting region, of the order of $10^4$
km. 
% The spectrum
% of the source in this interpretation resembles vaug to the
% peculiar spectrum shown by the peculiar Z-source Cir X--1 when it was 
% observed by BeppoSAX in
% a very high luminosity state (Iaria et al. 2001): it is therefore well
% possible that this spectral shape is a peculiar signature of accretion
% at very high accretion rates.
 It is possible that the emission from
the inner parts of the system are somehow 'obscured' by the matter
that is ejected from the system due to Eddington-limit accretion. Thus
the Comptonized component we see originates by the reemission of radiation
from the outer 
parts of the matter cloud that forms in the central zone of the disk. 
In this hypothesis, the system should be seen nearly
edge-on, as the central source is almost completely obscured by disk
matter, inflated in the central regions of the system.
 It is anyway hard to believe that the total
luminosity we get (about 1-2 times the Eddington luminosity) can come
from a region that far from the central compact object. 
Moreover, the system should be seen completely edge-on, but no
eclipse or X-ray dip has been
observed so far in the LMC X--2 lightcurve (Smale \& Kuulkers 2000).

The Birmingham model, is not able to fit the data satisfactorily, since 
the  component described by a cutoff power-law is incompatible with a
Comptonized spectrum (the power-law index is below 1), and if we fix
the value of the power-law index we cannot constrain the cutoff energy
due to the narrow band of XMM/Newton. 

Since all other models are physically unreasonable for one reason or
another,
the only model that fits satisfactorily the data consists of a
blackbody at 1.5 keV and a cooler disk blackbody at 0.8 keV (Eastern Model), that is the emission is dominated by the radiation
coming from the central source, plus a blackbody component due to the
thermal emission of the disk. The hot component coming from the compact object
(or from the boundary layer between the disc and the
compact object), in this case is well described by a
blackbody emission. This means that Comptonization of the boundary
layer emission is not detected in the emission of LMC X--2. This blackbody is
relatively hot, is the origin of most of the emission
from the source ($\sim 70 \%$ of the unabsorbed luminosity, see table 1), and has a radius that approaches the neutron star
radius: we find a blackbody radius $R_\infty =
15~\mathrm{km}$, which means that, after applying proper the general relativistic change of coordinates, the emission radius in the local coordinate system of the NS is $R_\mathrm{em} = R_\infty(1- GM/R_\infty c^2) \sim
13~\mathrm{km}$.
The cooler disk blackbody that describes the emission from the
accretion disc has an inner emission radius lower limit of $R_\infty\sim 30$
km, which is almost surely larger than the inner disc radius given the
extremely high luminosity (and thus accretion rate) of the source: the
inner disc radius for a source accreting at the Eddington limit is
typically at (or very near to) the surface of the NS. This 
means that we do not see any emission from the innermost parts of
the disc. This could be explained with the obscuration of that part of
the disc due to ejection of matter at the Eddington limit.
According to the classic paper on accretion disk structure by Shakura \& Sunyaev (1973) and to more recent works (see e.g. Miller \& Lamb 1993),
the inner parts of a disc at supercritical accretion rates is inflated by radiation pressure, that becomes dominant at the so-called
\textit{spherization} radius (Shakura \& Sunyaev 1973) $$r_{sph} \simeq 17.3 \dot{M}/10^{-8}~\mathrm{km},$$ that in this case in placed at $22$
km from the compact object.

 This model has been
widely used to describe the spectra of Z sources, although often the
emission of the central source is Comptonized (see e.g.  Di Salvo et al
2000, Di Salvo et al
2001, Di Salvo et al. 2002). This model of
the geometry of emission is often opposed to the Birmingham model, where the
emission is described by a blackbody coming from the boundary layer of the disk with the compact object, plus a comptonized component describing
the emission of an extended accretion disc corona, which is usually
fitted with a cutoff powerlaw, where the cutoff energy is equal to the
temperature of the scattering electrons cloud. We have shown that the
Birmingham model, which has been used to fit the low-resolution
spectra of several 
Z-sources (Sco X--1, Barnard et al. 2003; GX~340+0, Church et al. 2006),
 cannot give a good fit of the spectrum of LMC X--2. This could be
 due to the fact that at the high accretion rates that we have in LMC
 X--2, the thermal
 emission of the accretion disc dominates over any emission from an
 accretion disc corona.

\subsection{Discrete spectral features}

LMC~X--2 is the second, among the 8 Z-sources, which does not show any
emission feature in the Fe region (after GX~5-1, Asai et
al. 1994).We have shown that it is not likely that the non-detection could be due to the lack of statistics. The absence of this feature in the
spectrum could then be due to the lower metallicity of the stars in LMC
($Z\sim 0.008$) compared to the mean metallicity  
of the stars in our galaxy ($Z \sim 0.02$): if the donor is less
abundant in heavy 
elements, so is the disc and thus any emission line will be fainter as
well by roughly the same proportionality factor. 

\begin{figure}[h]

  \centering
  \includegraphics{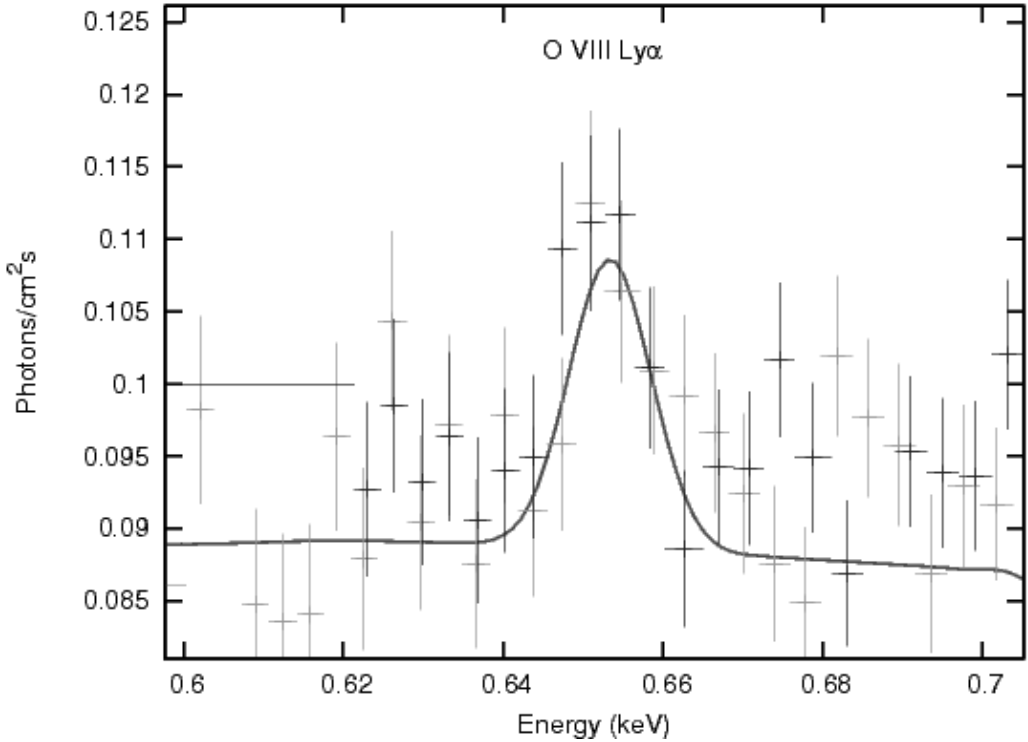}
  \caption{Unfolded spectrum in the region of the O VIII emission line. We show data (dark and light grey for RGS1 and RGS2, respectively) toghether with the total model (continuum + line).}
  \label{fig:oline}
\end{figure}
On the other hand, we detect  for the first time in a Z-source, a O VIII
Ly-$\alpha$ emission line with a significance of 4.4 $\sigma$ (see
figure \ref{fig:oline}). The energy of the emission line is perfectly
compatible with the rest-frame energy of the O VIII Ly $\alpha$
transition energy (653.4 eV).  

This emission line could be present in most
Z-sources as O is more abundant than heavier elements in the
donor star, but is normally undetected due to the strong absorption by
neutral matter that is present in all Galactic Z-sources, as they lie
relatively near the galactic center. In the case of LMC X--2, the
absorption due to neutral matter is more than an order of magnitude
lower than for Galactic sources, and this can explain why we do detect
this line, while we do not detect any other emission line at higher
energies probably due to the low metallicity of LMC stars.

\section{Conclusions}

We have analyzed an archival XMM/Newton observation of LMC X--2, which
allowed us to redetermine the position of the source
with respect to the value reported in the Rosat All Sky Survey
catalog. The source was probably in the normal or in the flaring
branch, as its timing characteristics and high luminosity
(exceeding the Eddington limit) exclude that the source is in the
horizontal branch. Analyzing the spectra of the 3 active instruments
we found that they could be well fitted by a two component model
(blackbody plus disk blackbody) plus an emission line at 653 eV, which
is identified as an O VIII Ly-$\alpha$ emission line. On the contrary, the iron
region lacks any emission line. As discussed, this can be caused by
the low metallicity of the stars in the LMC. 
\acknowledgements{We thank the referee for useful suggestions that improved our paper. This work was supported by the Ministero della Istruzione, della 
Universit\`a e della Ricerca (MIUR) and by the Agenzia Spaziale Italiana (ASI).}

\end{document}